\documentclass[11pt,a4paper]{article}
\usepackage[a4paper,margin=3cm]{geometry}
\usepackage[colorlinks=true
,urlcolor=blue
,anchorcolor=blue
,citecolor=blue
,filecolor=blue
,linkcolor=blue
,menucolor=blue
,linktocpage=true
,pdfproducer=medialab
,pdfa=true
]{hyperref}

\usepackage{multirow}
\usepackage{tikz}
\usepackage[super]{nth}
\usepackage{amsmath}
\usepackage{graphicx}
\usepackage{titlesec}
\usepackage{epstopdf}
\usepackage{needspace}

\normalsize
\newcommand\acknowledgments{\section*{Acknowledgments}}

\newcommand{\GeV}[1]{#1\,\mbox{GeV}}
\newcommand{\TeV}[1]{#1\,\mbox{TeV}}
\newcommand{\invfb}[1]{#1\,\mbox{fb}^{-1}}
\newcommand{\invab}[1]{#1\,\mbox{ab}^{-1}}

\newcommand{\BR}[1]{\mathcal{B}(#1)}
\newcommand{\nsigma}[1]{#1\,\sigma}

\graphicspath{{figures/}}

\title{PSI/UZH Workshop:\\
	Impact of $B\to\mu^+\mu^-$ on \\[0.5em]New Physics Searches}
\author{Andreas Crivellin, Margherita Ghezzi, Dario M\"uller, Adrian Signer and Yannick Ulrich}
\date{\nth{18}-\nth{19} December 2017 at the Paul Scherrer Institute}

\titlespacing*{\subsection}{0pt}{5em}{0.7em}
\def\abbrauthordo#1#2 #3\relax{#1. #3}

\newcommand{\talk}[4]{
    \needspace{5cm}
    \subsection{#1}
    \vspace{-3mm}
    \begin{flushright}\textit{Speaker: #2}\\#3\end{flushright}
    \vspace{-3mm}
    \IfFileExists{talks/#4.tex}{\nopagebreak\input{talks/#4}}{to be written}
}
\makeatletter
\g@addto@macro\bfseries{\boldmath}

\newcommand{\preprint}[1]{\def\@preprint{#1}}
\renewcommand{\abstract}[1]{\def\@abstract{#1}}
\def\@maketitle{
\ifdefined\@preprint{
    \begin{flushright}
        \parbox[][5cm][t]{3cm}{\raggedleft\@preprint}
    \end{flushright}\vspace{-2cm}}\fi
\raggedright
\vspace{-2cm}
\begin{center}
{\Huge \bfseries \sffamily \@title }\\[4ex] 
\@date\\[8ex]
\includegraphics[width=0.6\textwidth]{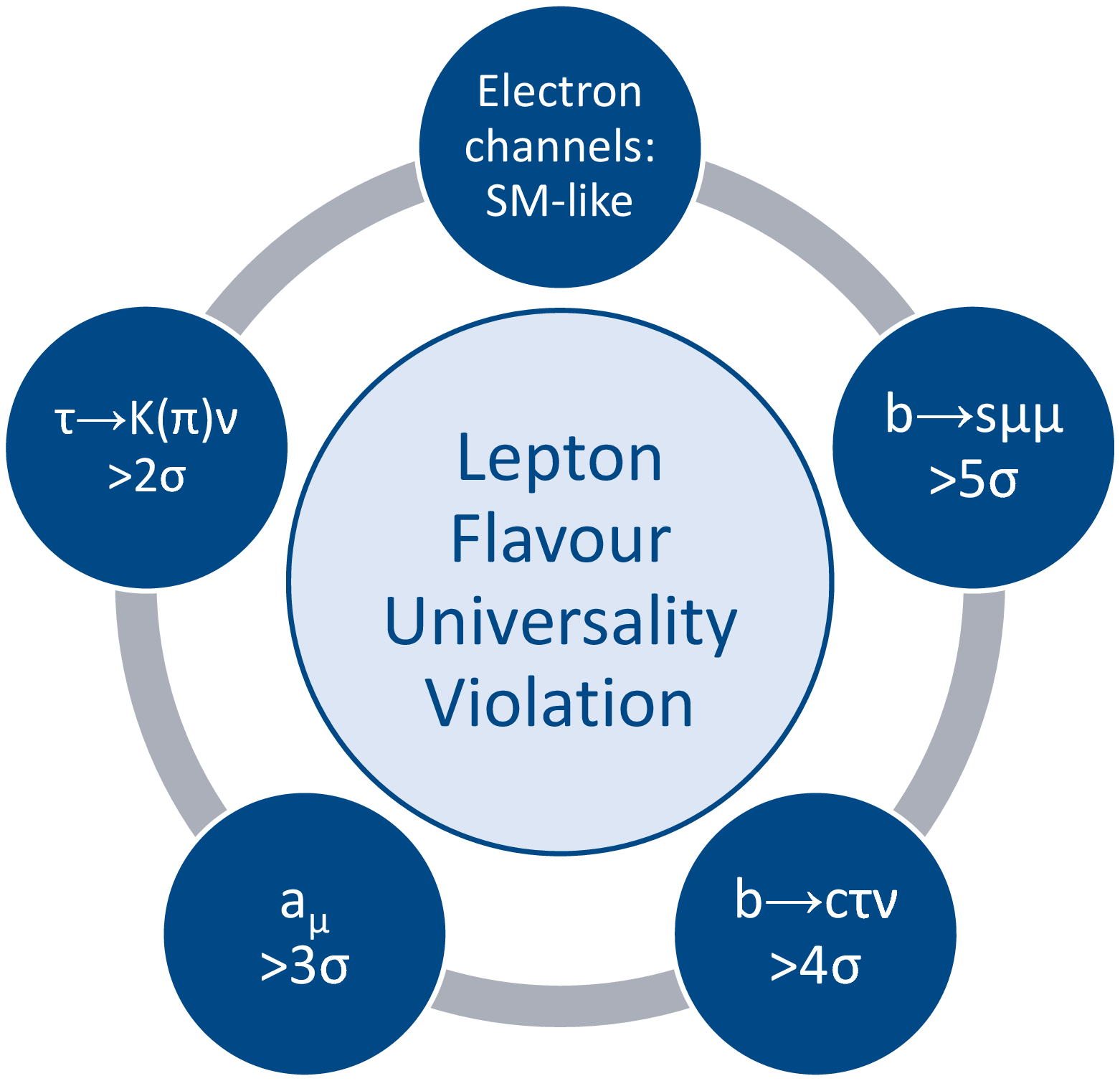}
\end{center}

\vspace{0.5cm}

\textbf{Speakers}: F.~Bernlochner,  A.~Crivellin, I.~de Medeiros Varzielas,  S.~Descotes-Genon,  M.~Fael,  D.~Ghosh,  A.~Greljo,  M.~Hoferichter,  G.~Isidori,  U.~Langenegger,  M.~Misiak,  M.~Mulder,  U.~Nierste,  A.~Papa,  M.~Rama,  P.~Reznicek and  D.~Straub
\begin{center}

\vspace{0.5cm}

\textbf{Abstract}\\[2ex]
\begin{minipage}[]{0.9\textwidth} {} 
\ifdefined\@abstract{\@abstract}\fi
\end{minipage}
\end{center}
\vspace\fill
\textbf{Editors}: \@author\\[2ex]
}
\makeatother

\preprint{PSI-18-05\\ZH-TH-11/18}
\abstract{In these mini-proceedings we review the results of the
workshop ``Impact of $B\to\mu^+\mu^-$ on New Physics Searches'' that
took place at the Paul Scherrer Institute (PSI) on the
\nth{18}-\nth{19} December 2017.}

\begin{document}
\maketitle

\clearpage

\section{Introduction}

So far, the LHC has not directly observed any particle beyond the ones
of the Standard Model (SM). However, in recent years interesting hints
for New Physics (NP) have been accumulated in flavour physics.

These discrepancies with the SM predictions are most pronounced in
semi-leptonic $B$ decays.  Here, we have two classes of processes:
\begin{itemize}
	\item
    $b\to c\tau\nu$: In these processes, mediated at tree-level in the
    SM, several measurements like
    \begin{align}\begin{split}
        R_\tau(X) &\equiv \frac{\BR{B\to X\,\tau\nu_\tau}}{\BR{B\to
        X\, \ell\nu_\ell}}
            \qquad\text{with } X=D,\,D^*
        \,,\\
        R_\tau(J/\psi) &\equiv \frac{\BR{B_c\to
        J/\psi\,\tau\nu_\tau}}{\BR{B_c\to J/\psi\, \ell\nu_\ell}}
    \end{split}\end{align}
    with $\ell=e,\mu$ point towards lepton flavour universality violation
    (LFUV) in $\tau-\mu,e$ at the $\approx \nsigma{4}$
    level~\cite{Amhis:2016xyh}.
    
    \item 
    $b\to s\ell^+\ell^-$: This flavour changing neutral current
    process is loop suppressed and is proportional to the CKM element
    $V_{ts}$.  Here the measurements of $R_\mu(K)$~\cite{Aaij:2014ora}
    and $R_\mu(K^{*})$~\cite{Aaij:2017vbb}, defined as
    \begin{align}
        \hspace{-3.105cm}
        R_\mu(X) \equiv \frac{\BR{B\to X\mu^+\mu^-}}{\BR{B\to X e^+e^-}}
        \,,
    \end{align}
    are supported by other $b\to s\mu^+\mu^-$ observables (like
    $P_5^{\prime\mu}\equiv P_5^\prime$ as defined
    in~\cite{Aaij:2015oid}) which also show deviations from the SM
    predictions.
\end{itemize}

In the second class of processes, the decay $B_s\to\mu^+\mu^-$ is
included. Currently, the measurements are in agreement with the SM.
However, while the theory predictions are quite precise, the
experimental errors are still as large as the effect one can expect
from other $b\to s\mu^+\mu^-$ observables like $R_\mu(K^{(*)})$,
$P_5^\prime$, etc. Nonetheless, in the future we can expect
significant progress in $B_s\to\mu^+\mu^-$ and this decay will play a
key role in distinguishing among different NP scenarios. Furthermore,
with increasing statistics, one can also improve the search for $b\to
d\ell^+\ell^-$ transitions, closing in on the SM predictions and
making $B_d\to\mu^+\mu^-$ a golden mode for a NP discovery.

In this workshop, we discussed the current experimental status of
$b\to s\ell^+\ell^-$ transitions (with focus on $B\to\mu^+\mu^-$) as
well as possible connections to $b\to c\tau\nu$, charged lepton
flavour violation (LFV), and kaon physics.

\section{Summary of presented talks}
In the following sections we will summarise the talks presented during
the workshop. The speakers' slides are available on the conference web
page
{\small\href{http://indico.cern.ch/event/655338/}{\texttt{indico.cern.ch/event/655338/}}}.

    \needspace{5cm}
    \subsection{\texorpdfstring{$B\to\mu^+\mu^-$}{B->mu mu} at CMS}
    \vspace{-3mm}
    \begin{flushright}\textit{Speaker: Urs Langenegger}\\Paul Scherrer Institute\end{flushright}
    \vspace{-3mm}
    \nopagebreak CMS searched for both $B_s\to\mu^+\mu^-$ and $B_d\to\mu^+\mu^-$
events in the Run-1 data, corresponding to $\invfb{25}$. The current
status of these measurements, using an unbinned maximum likelihood
fit, is~\cite{Chatrchyan:2013bka}
\begin{align}\begin{split}
\BR{B_s\to\mu^+\mu^-}_{\rm CMS}&=\left(3.0^{+1.0}_{-0.9}\right)\times 10^{-9}\,,\\
\BR{B_d\to\mu^+\mu^-}_{\rm CMS}&<1.1\times10^{-9}\quad(95\%\,\mathrm{C.L.})\,.
\end{split}\end{align}
The $B_s\to\mu^+\mu^-$ signal yield is incompatible with the
background-only hypothesis at $\nsigma{4.3}$. In addition, CMS also
performed an analysis of $B\to K^*\mu^+\mu^-$ determining $P_1$ and
$P_5^\prime$~\cite{CMS:2017ivg}. Also here, the results are compatible
with the SM predictions but still have quite large errors due to
statistics.

By 2017 the integrated luminosity has been increased by a factor of three
and the Run-2 analysis is pursued with high priority. An analysis with
$\invfb{300}$ will give a 12\% accuracy in $ \BR{B_s\to\mu^+\mu^-}$
and a 47\% accuracy in
$\BR{B_d\to\mu^+\mu^-}$~\cite{Sirunyan:2017dhj}. Furthermore, an
analysis of $R_\mu(K)$, testing the current LHCb results, is forthcoming.

    \needspace{5cm}
    \subsection{\texorpdfstring{$B\to \ell^+\ell^-$}{B-> ll} at LHCb}
    \vspace{-3mm}
    \begin{flushright}\textit{Speaker: Matteo Rama}\\INFN Pisa\end{flushright}
    \vspace{-3mm}
    \nopagebreak A new LHCb branching fraction measurement in 2017 uses an improved
analysis as well as more data. The results are~\cite{Aaij:2017vad}
\begin{align}
\begin{split}
\BR{B_{s}\to\mu^{+}\mu^{-}}_{\rm LHCb}&=\left(3.0\pm0.6^{+0.3}_{-0.2}\right)\times10^{-9}\,,\\
\BR{B_{d}\to\mu^{+}\mu^{-}}_{\rm LHCb}&=\left(1.5^{+1.2+0.2}_{-1.0-0.1}\right)\times 10^{-10}\\
            &<3.4\times10^{-10}\quad(95\%\,\mathrm{C.L.})\,.
\end{split}
\end{align}
The latter branching ratio disfavours the background-only hypothesis
at $\nsigma{1.6}$. Both results are compatible with the SM predictions.

In addition, the $B_{s}\to\mu^{+}\mu^{-}$ effective lifetime was
measured for the first time \cite{Aaij:2017vad}:
\begin{align}
\tau\left(B_{s}\to\mu^{+}\mu^{-}\right)_{\rm LHCb}
        =\Big(2.04\pm0.44\pm0.05\Big) {\rm ps}\,.
\end{align}
This is consistent with the lifetime asymmetry
$A_{\Delta\Gamma}^{\mu^{+}\mu^{-}}=1(-1)$ hypothesis at
$\nsigma{1.0(1.4)}$.

In addition, LHCb also searched for $B_{q}\to\tau^{+}\tau^{-}$
decays, which are experimentally very challenging. The $B_{s}$ and
$B_{d}$ decays cannot be separated and hence, assumptions on one decay
are needed to extract the limit on the other one. LHCb
obtains~\cite{Aaij:2017xqt}
\begin{align}
 \refstepcounter{equation}
\BR{B_{s}\to\tau^{+}\tau^{-}}_{\rm LHCb}<5.2(6.8)\times10^{-3}
\quad(90(95)\%\,\mathrm{C.L.})\,,\tag{\theequation a}
\end{align}
under the assumption that the signal is fully dominated by $B_{s}$. On
the other hand, assuming that it is fully dominated by $B_{d}$, one
finds
\begin{align}
\BR{B_{d}\to\tau^{+}\tau^{-}}_{\rm LHCb}<1.6(2.1)\times10^{-3}
\quad(90(95)\%\,\mathrm{C.L.})\,.\tag{\theequation b}
\end{align}
There is also a recent update on the LFV decays $B\to
e^{\pm}\mu^{\mp}$~\cite{Aaij:2017cza}:
\begin{align}\begin{split}
\BR{B_{s}\to e^{\pm}\mu^{\mp}}_{\rm LHCb}<5.4(6.3)\times10^{-9}
\quad(90(95)\%\,\mathrm{C.L.})\,,\\
\BR{B_{d}\to e^{\pm}\mu^{\mp}}_{\rm LHCb}<1.0(1.3)\times10^{-9}
\quad(90(95)\%\,\mathrm{C.L.})\,,
\end{split}\end{align}
which is a factor 2-3 more precise than the previous LHCb measurements.

    \needspace{5cm}
    \subsection{\texorpdfstring{$b\to s\ell^+\ell^-$}{b -> sll} at LHCb}
    \vspace{-3mm}
    \begin{flushright}\textit{Speaker: Mick Mulder}\\Nikhef\end{flushright}
    \vspace{-3mm}
    \nopagebreak

The most significant deviations from the SM predictions in $b\to
s\ell^+\ell^-$ processes were found by the LHCb collaboration. In
2013, using the $\invfb1$ dataset, the LHCb experiment measured the
optimised observables of~\cite{Descotes-Genon:2013vna} for $B\to
K^*\mu^+\mu^-$~\cite{Aaij:2013qta}, observing  a sizeable
$\nsigma{3.7}$ discrepancy in one bin of $P_5^\prime$. This was later
confirmed by the analysis of the $\invfb3$ dataset, finding $\nsigma3$
deviations in each of two adjacent bins at large $K^*$
recoil~\cite{Aaij:2015oid}. Furthermore, LHCb also measured the
branching ratios of several semi-leptonic $B$
decays~\cite{Aaij:2013aln,Aaij:2015esa} observing a systematic deficit
with respect to SM predictions. 

\begin{figure*}[b]
	\begin{center}
		\begin{tabular}{cp{7mm}c}
			\includegraphics[width=0.45\textwidth]{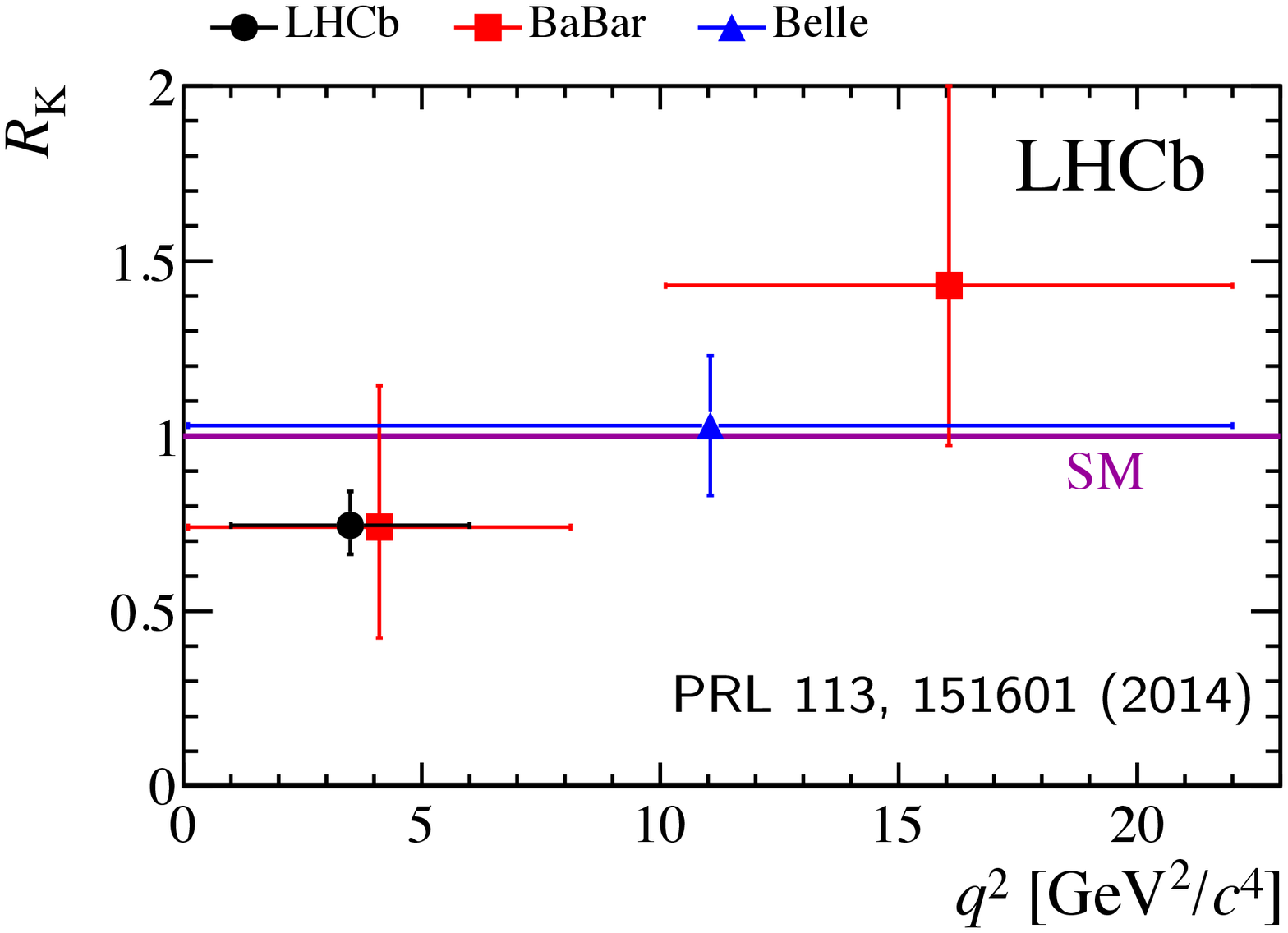}
			\includegraphics[width=0.45\textwidth]{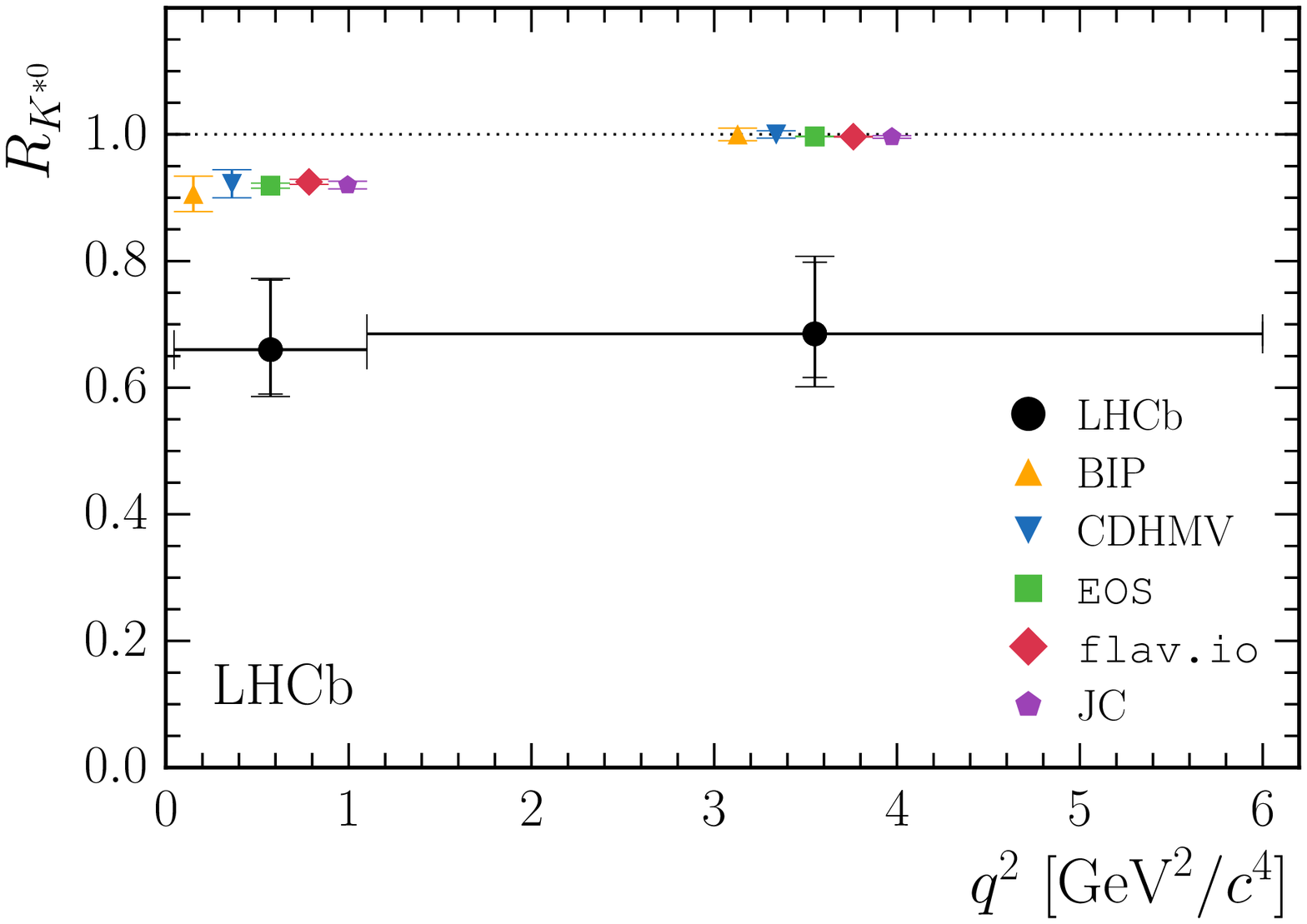}
		\end{tabular}
	\end{center}
	\caption{Measurements and SM predictions of $R_\mu(K)$ and
    $R_\mu(K^*)$. Figure taken from~\cite{Aaij:2014ora,Aaij:2017vbb}.} 
	\label{RKRKs}
\end{figure*}

A conceptually new element arose when the ratios
$R_\mu(K)$~\cite{Aaij:2014ora} and $R_\mu(K^*)$~\cite{Aaij:2017vbb}
were measured and found to be significantly below the SM predictions
which are very close to one (see Figure~\ref{RKRKs}).

Currently, many more tests of LFUV, such as $R_\mu(\phi),\,
R_\mu(\Lambda)$ or $P_5^{\prime \mu}-P_5^{\prime e}$, are under
investigation at LHCb and significant improvements
on LFV decays like $B\to K \mu^\pm e^\mp$ or $\Lambda_b^0\to \Lambda^0
\mu^\pm e^\mp$ are expected.

    \needspace{5cm}
    \subsection{\texorpdfstring{$b\to s\ell^+\ell^-$}{b -> sll} at ATLAS}
    \vspace{-3mm}
    \begin{flushright}\textit{Speaker: Pavel Reznicek}\\Charles University\end{flushright}
    \vspace{-3mm}
    \nopagebreak

The ATLAS experiment performed measurements of
$B\to\mu^+\mu^-$~\cite{Aaboud:2016ire} and an angular analysis of
$B_d\to K^{*0}\mu^+\mu^-$~\cite{ATLAS:2017dlm}.

The $B\to\mu^+\mu^-$ analysis was done using the full Run-1 dataset
of $\invfb{(4.9+20)}$ at 7 and $\TeV8$ respectively with a muon
threshold of $p_T(\mu) > 4\,\GeV$. The branching ratio measurement is
normalised to $B^\pm\to J/\psi (\mu^+\mu^-) K^\pm$ and corrected by
Monte Carlo to account for the different production cross sections and
the different detector acceptances and efficiencies.  To discriminate
signal from background, a boosted decision tree (BDT) was trained on
Monte Carlo data and tested on the data side bands. The result is
compatible with the SM at $\nsigma{2.0}$
\begin{align}\begin{split}
\BR{B_s\to\mu^+\mu^-}_{\rm ATLAS}&=\left(0.9^{+1.1}_{-0.8}\right)\times 10^{-9}\,,\\
\BR{B_d\to\mu^+\mu^-}_{\rm ATLAS}&<      4.2                     \times 10^{-10}
\quad(95\%\,\mathrm{C.L.})\,.
\end{split}\end{align}

In the most recent analysis of $B_d\to K^{*0}\mu^+\mu^-$ using
$\invfb{20.3}$ of $\TeV8$ data the decay angles were fitted in the
most general parametrisation allowed by the process. The analysis,
which was done up to $q^2=m_{\mu^+\mu^-}^2 < \GeV6$, suffers mostly
from its low statistics but allows one to cancel large hadronic
uncertainties from the form factors by defining the $P_i^{(\prime)}$
variables.
The results are roughly compatible with the SM predictions, with the
largest deviation of $\nsigma{2.5}$ in $P_5'$ which is compatible with
the LHCb measurement.

In the future, detector upgrades and new trigger strategies will help
to cope with the higher luminosity. Further, the feasibility of
measuring final states including electrons is under investigation.

    \needspace{5cm}
    \subsection{\texorpdfstring{$B\to\mu^+\mu^-$}{B->mu mu} theory status}
    \vspace{-3mm}
    \begin{flushright}\textit{Speaker: Mikolaj Misiak}\\University of Warsaw\end{flushright}
    \vspace{-3mm}
    \nopagebreak The average time-integrated branching ratios for $B_q \to\ell^+\ell^-$ within
the SM are known very precisely. They are given
by~\cite{Bobeth:2013uxa}
\begin{equation}
\overline{\mathcal{B}}_{q\ell} = \left| V_{tb} V_{tq}^* \right|^2 G_F^4\, m_W^4\, 
\frac{ m_\ell^2\, m_{B_q}\, f_{B_q}^2 }{ 2\pi^5\Gamma_H^q }
\sqrt {1 - \frac{ 4m_\ell^2 }{ m_{B_q}^2}}\, \eta_{\scriptscriptstyle\rm QED} \left|C_A(\mu_b) \right|^2\,,\label{BmumuBrSM}
\end{equation}
where all the masses are assumed to be renormalised on shell, and $\Gamma_H^q$
is the decay width of the heavier eigenstate in the $B_q$-$\bar B_q$
system. Perturbative calculations of the Wilson coefficient $C_A$, including
the three-loop QCD~\cite{Hermann:2013kca} and two-loop electroweak
corrections~\cite{Bobeth:2013tba}, give 
$C_A(\mu_b) \simeq 0.4690 \left(M_t/\GeV{173.1}\right)^{1.53}
                          \left(\alpha_s(m_Z)/0.1184\right)^{-0.09}$ 
at $\mu_b = \GeV{5}$~\cite{Bobeth:2013uxa}. In the global $b \to s\ell^+\ell^-$
fits, $C_{10} = -2 C_A/\sin^2\theta_W$ can be used instead, after carefully
adjusting the renormalization scheme for $\sin^2\theta_W$.

All the non-perturbative inputs in Eq.~\eqref{BmumuBrSM} that come from
theoretical calculations are encoded in the $B_q$-meson decay constants
$f_{B_q}$ (that can reliably be calculated using lattice QCD) as well as in
the QED correction factor $\eta_{\scriptscriptstyle\rm QED}$. The latter
factor stands for QED effects emerging below the scale $\mu_b$, and depending
on several Wilson coefficients.  A recent calculation~\cite{Beneke:2017vpq} of
power-enhanced QED effects gives $\eta_{\scriptscriptstyle\rm QED} = 0.993 \pm
0.004$.

Using the inputs from Table~\ref{tab:bmm.update}, we obtain the SM predictions
$\overline{\mathcal{B}}_{s\mu} = (3.54 \pm 0.21)\times 10^{-9}$ 
and
$\overline{\mathcal{B}}_{d\mu} = (1.00 \pm 0.07)\times 10^{-10}$.
Apart from the parametric uncertainties, an additional uncertainty
of $1.5\%$ has been included in the above predictions, following the 
estimate in~\cite{Bobeth:2013uxa}.

The largest uncertainty at the moment is parametric, coming mainly from the
decay constants $f_{B_q}$ and CKM elements $V_{tq}$. In the $B_s \to \ell^+
\ell^-$ case, the element $V_{ts}$ is linked via unitarity to
$V_{cb}$. However, the determination of $V_{cb}$ suffers from tensions between
the inclusive and exclusive calculations. Since these tensions cannot be
attributed to NP~\cite{Crivellin:2014zpa}, it is very important to improve on
the determinations of $V_{cb}$ to reduce the uncertainty in
$\overline{\mathcal{B}}_{s\ell}$. Here, only the inclusive
determination~\cite{Gambino:2016jkc} of $V_{cb}$ has been used.

\begin{figure}[t]
\begin{center}\small
\begin{tabular}{c|l|l}
parameter                    & value       & source \\\hline
&&\\[-5mm]
$M_t\,[\mbox{GeV}]$          & 174.30(65)  & \cite{TevatronElectroweakWorkingGroup:2016lid}\\[1mm]
$\alpha_s(m_Z)$              & 0.1182(12)  & \cite{Patrignani:2016xqp}\\[1mm]
$f_{B_s}\,[\mbox{GeV}]$      & 0.2240(50)  & \cite{Aoki:2016frl,Dowdall:2013tga}\\[1mm]
$f_{B_d}\,[\mbox{GeV}]$      & 0.1860(40)  & \cite{Aoki:2016frl,Dowdall:2013tga}\\[1mm]
$|V_{cb}|$                   & 0.04200(64) & \cite{Gambino:2016jkc}\\[1mm]
$|V^*_{tb} V_{ts}|/|V_{cb}|$ & 0.9819(4)   & derived from~\cite{Charles:2015gya}\\[1mm]
$|V^*_{tb} V_{td}|$          & 0.0087(2)   & derived from~\cite{Charles:2015gya}\\[1mm]
$1/\Gamma_H^s\,$[ps]         & 1.619(9)    & \cite{Amhis:2016xyh}\\[1mm]
$1/\Gamma_H^d\,$[ps]         & 1.518(4)    & \cite{Amhis:2016xyh}
\end{tabular}
\renewcommand{\figurename}{Table}
\caption{Input parameters used in evaluating the SM predictions for 
$\overline{\mathcal{B}}_{s\mu}$ and $\overline{\mathcal{B}}_{d\mu}$.}
\label{tab:bmm.update}
\end{center}
\end{figure}

    \needspace{5cm}
    \subsection{\texorpdfstring{$b\to s\mu^+\mu^-$}{b->s mu mu} global analysis}
    \vspace{-3mm}
    \begin{flushright}\textit{Speaker: S\'ebastien Descotes-Genon}\\Laboratoire de Physique Th\'eorique d'Orsay, CNRS\end{flushright}
    \vspace{-3mm}
    \nopagebreak A global six-dimensional fit to all available $b\to s\ell^+\ell^-$
data for the Wilson coefficients
$C_7^{(\prime)},\,C_9^{(\prime)},C_{10}^{(\prime)}$ confirms the need
for a large contribution to $C_9$, with a SM pull reaching
$\nsigma{5.0}$ once the $R_\mu(K^*)$ measurement is
included~\cite{Capdevila:2017bsm}. Hadronic uncertainties conform to
theoretical expectations~\cite{Capdevila:2017ert} and unexpectedly
large effects (power corrections to form factors, charm-loop
contributions) are disfavoured by the significant amount of LFUV
observed. 

While a sizeable effect in $C_9$ with muons is required
($\mathcal{O}(25\%)$ of the SM contribution), an effect in $C_{10}$,
including even $C_9=-C_{10}$, is possible. Even though $R_\mu(K)$ and
$R_\mu(K^*)$ can already tell us about NP effects in $C_{9}$,
$B_s\to\mu^+\mu^-$ is crucial for determining if $C_{10}$ is SM-like
or not. Therefore, if $C_9=-C_{10}$ is realised in nature, one expects
a significant reduction of $\BR{B_s\to\mu^+\mu^-}$ compared to
the SM.
 
\begin{figure*}[t]
	\begin{center}
        \hspace{1.7cm}\includegraphics[width=0.65\textwidth]{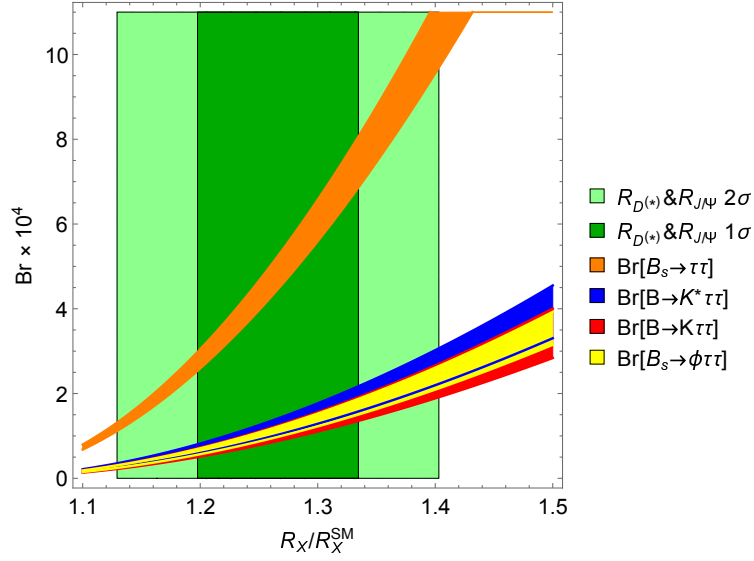}
	\end{center}
	\caption{Predictions of the branching ratios of the $b\to
    s\tau^+\tau^-$ processes (including uncertainties) as a function
    of $R_\tau(X)/R_\tau(X)^{\rm SM}$. Figure taken
    from~\cite{Capdevila:2017iqn}.}         
	\label{fig:predsbstautau}
\end{figure*}

Another possibility to establish the presence of NP in $b\to
s\ell^+\ell^-$ transitions is looking at $b\to s\tau^+\tau^-$
processes. Here, $R_\tau(D)$, $R_\tau(D^*)$ and $R_\tau(J/\psi)$
suggest that $b\to s\tau^+\tau^-$ processes can be enhanced by up to
three orders of magnitude compared to the SM~\cite{Capdevila:2017iqn},
(see Figure~\ref{fig:predsbstautau}).

    \needspace{5cm}
    \subsection{\texorpdfstring{$b\to s\mu^+\mu^-$ and $b\to c\tau\nu$}{b->s mu mu and b->c tt} in the SM EFT}
    \vspace{-3mm}
    \begin{flushright}\textit{Speaker: Matteo Fael}\\University of Siegen\end{flushright}
    \vspace{-3mm}
    \nopagebreak In order to evaluate the predictions of the SM effective field theory (EFT),
realised above the EW breaking scale, the matching of the gauge
invariant dimension-six operators on the $B$ physics Hamiltonian
(including LFV operators) integrating out the top,
$W$, $Z$ and the Higgs has to be performed~\cite{Alonso:2014csa}.
Here, operators within the SMEFT involving top quarks are of special
interest since they do not contribute to $b\to s$ processes at the
tree level, because the top is not a dynamical degree of freedom of
the $B$ physics Hamiltonian. Therefore, all operators involving
right-handed top quarks can give numerically important contributions
at the one loop-level~\cite{Aebischer:2015fzz}: 
\begin{enumerate}
	\item 4-fermion operators to 4-fermion operators ($\Delta B =\Delta S =1$)
	\item 4-fermion operators to 4-fermion operators ($\Delta B =\Delta S =2$)
	\item 4-fermion operators to $O_7$ and $O_8$
	\item Right-handed $Z$ couplings to $O_9$, $O_{10}$ and $O^q_{3-6}$
	\item Right-handed $W$ couplings to $O_7$ and $O_8$
	\item Magnetic operators to $O_7$, $O_8$, $O_9$, $O_{10}$ and $O_4^q$
\end{enumerate}
In fact, for a vector operator with right-handed top quarks and
left-handed muons, this effect can be used to explain the anomalies in
$b\to s\mu^+\mu^-$ in UV complete models, either with a
$Z^\prime$~\cite{Kamenik:2017tnu} or with
leptoquarks~(LQ)~\cite{Becirevic:2017jtw}.

    \needspace{5cm}
    \subsection{Impact of future \texorpdfstring{$B\to\mu^+\mu^-$}{B->mu mu}}
    \vspace{-3mm}
    \begin{flushright}\textit{Speaker: David Straub}\\TU Munich\end{flushright}
    \vspace{-3mm}
    \nopagebreak

The $B_s \rightarrow \mu^+ \mu^-$ decay is particularly sensitive to
scalar NP contributions that can be parametrised in a
model-independent way by means of two complex parameters $S$ and
$P$~\cite{Altmannshofer:2017wqy}:
\begin{equation}
\BR{B_s \rightarrow \mu^+ \mu^-} = \frac{G_F^2 \alpha^2}{16 \pi^3} |V_{ts} V^*_{tb}|^2 f_{B_s}^2 \tau_{B_s} m_{B_s} m_\mu^2 \sqrt{1-\frac{4m_\mu^2}{m_{B_s}^2}} |C_{10}^{\mbox{\scriptsize SM}}|^2 \left( |P|^2 +|S|^2 \right)\,.
\end{equation}
Within an EFT approach, the effective operators giving rise to $S$ and
$P$ are 
\begin{align}\begin{split}
\mathcal O_{10}^{(\prime)} &=     \left(\bar s_{L(R)} \gamma_\rho b_{L(R)}\right)\left(\bar\mu\gamma^\rho\gamma^5\mu\right)\,,\\
\mathcal O_S   ^{(\prime)} &= m_b \left(\bar s_{L(R)} \gamma_\rho b_{R(L)}\right)\left(\bar\mu                   \mu\right)\,,\\
\mathcal O_P   ^{(\prime)} &= m_b \left(\bar s_{L(R)} \gamma_\rho b_{R(L)}\right)\left(\bar\mu\gamma^5           \mu\right)\,.
\end{split}\end{align}
The pseudoscalar contribution $P$ can be expressed in terms of the Wilson
coefficient combinations $C_{10}-C'_{10}$ and $C_P-C'_P$, while the scalar
contribution $S$ is proportional to $C_S-C'_S$. This leads to a degeneracy in
the parameter space that can be broken by the measurement of the
lifetime asymmetry $A_{\Delta\Gamma}^{\mu^+\mu^-}$ (up to a two-fold ambiguity)
\begin{equation}
A_{\Delta\Gamma}^{\mu^+\mu^-} = \frac{|P|^2
\cos\left(2\varphi_P-\varphi_s^{\mbox{\scriptsize NP}}\right) - |S|^2
\cos\left(2\varphi_S -\varphi_s^{\mbox{\scriptsize NP}} \right)}{|P|^2
+ |S|^2}\,.
\end{equation}

\begin{figure}[t]
	\begin{center}
		\begin{tabular}{cp{7mm}c}
			\includegraphics[width=0.45\textwidth]{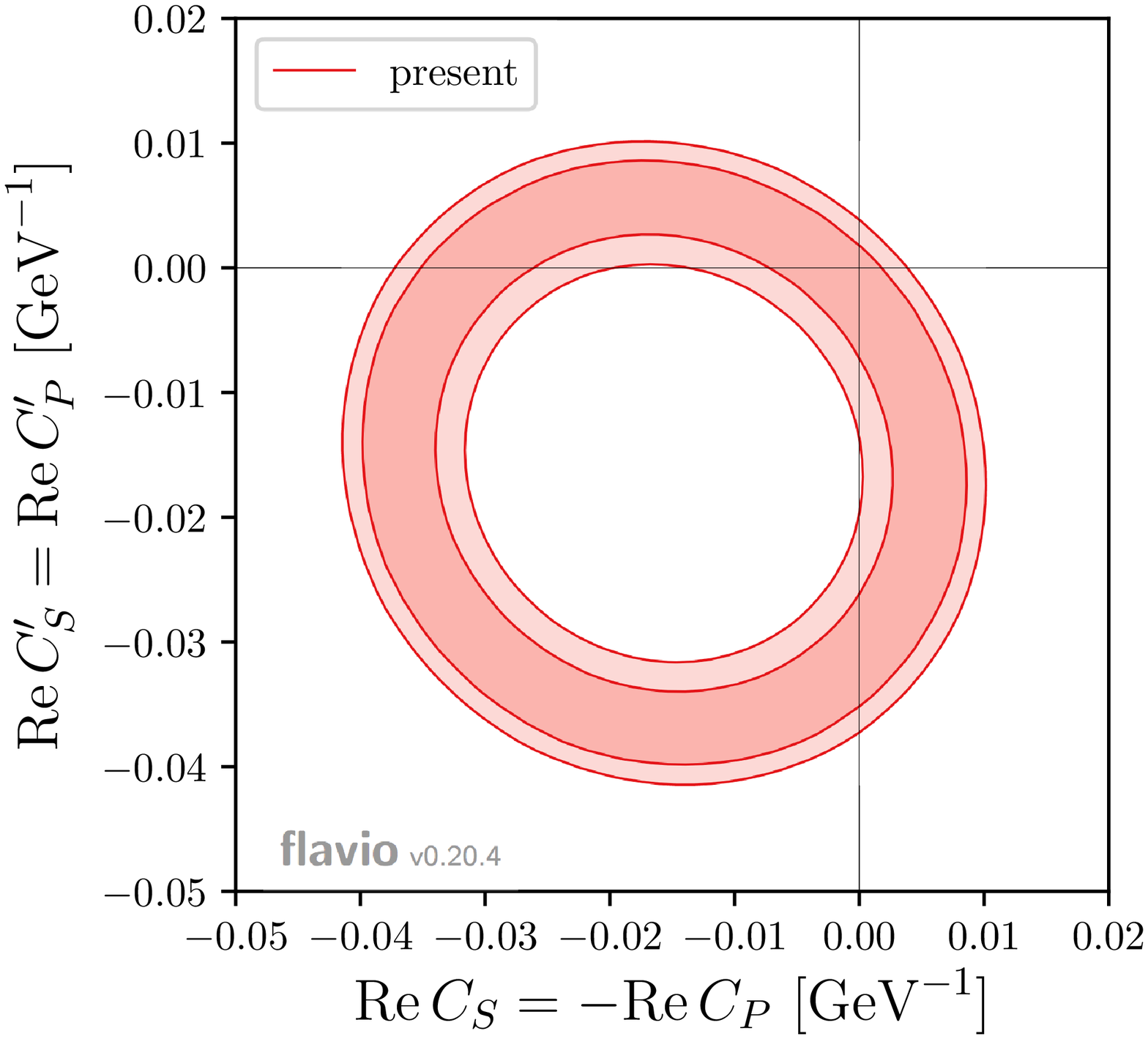}
			\includegraphics[width=0.45\textwidth]{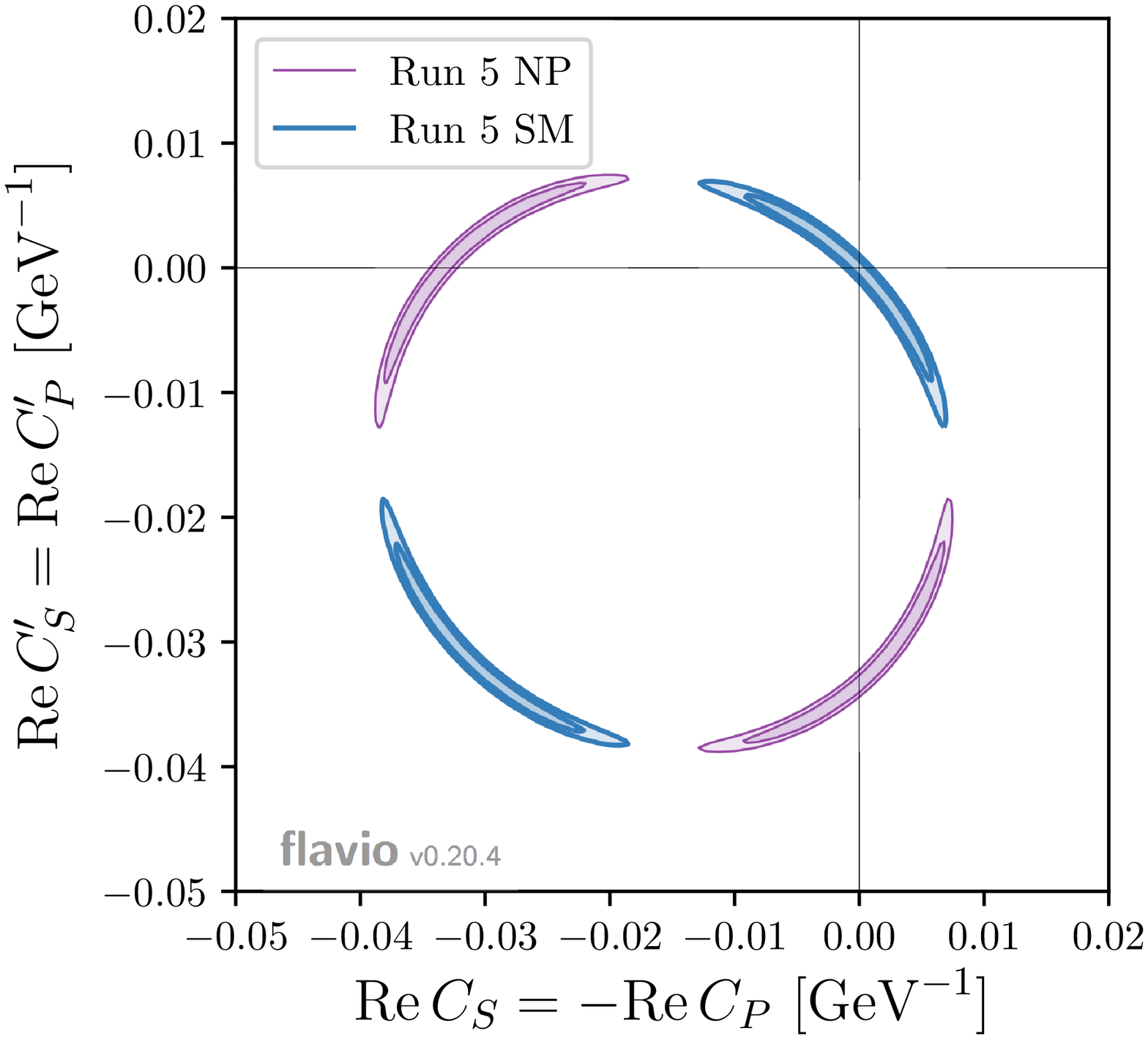}
		\end{tabular}
	\end{center}
	\caption{Present and future constraints on the real parts of the
    Wilson coefficients $C_S$ and $C'_S$ ($\nsigma1$ and $\nsigma2$ contours). Figure taken from~\cite{Altmannshofer:2017wqy}.}   
	\label{CS_CSp}
\end{figure}

Large effects in the coefficients $C_{S,P}^{(')}$ arise at tree level
mediated by a neutral scalar $\varphi \sim (1,2,-1)$ or vector LQ
in the representations $U_1 \sim (\bar{3},1,-4/3)$ or $V_2 \sim (
3,2,-5/3)$. In other contexts, like the MFV MSSM, these contributions
arise at the loop level. In Figure~\ref{CS_CSp} the current constraints
on the Wilson coefficients $C_S$ and $C'_S$ are shown, together with
the expected constraints at the future run 5.

Constraints from the $B_s \rightarrow \mu^+ \mu^-$ branching ratio and
the lifetime asymmetry are shown to be complementary to direct
searches in probing NP.

    \needspace{5cm}
    \subsection{Colloquium: Precision studies in flavour physics: a gateway to new laws of nature}
    \vspace{-3mm}
    \begin{flushright}\textit{Speaker: Ulrich Nierste}\\Karlsruhe Institute of Technology\end{flushright}
    \vspace{-3mm}
    \nopagebreak Even though not directly related to the hints for LFUV, there are also
interesting hints for NP in $\epsilon^\prime/\epsilon$ at the
$\nsigma3$ level~\cite{Buras:2015jaq}. While neither the tensions in
$b\to s\mu^+\mu^-$~\cite{Altmannshofer:2012az} nor in $b\to
c\tau\nu$~\cite{Fajfer:2012jt} can be explained by the most common
models of NP, the plain MSSM can account for
$\epsilon^\prime/\epsilon$~\cite{Kitahara:2016otd}. This is possible
due to cancellations between crossed and uncrossed box-diagram
contributions to kaon mixing~\cite{Crivellin:2010ys}. In this case,
interesting correlations with $K\to\pi\nu\bar\nu$
arise~\cite{Crivellin:2017gks} which can be tested at NA62.

    \needspace{5cm}
    \subsection{Experimental overview on \texorpdfstring{$R_\tau(D)$ and $R_\tau(D^*)$}{R(D) and R(D*)}}
\begin{figure*}[t]
	\begin{center}
        \includegraphics[width=0.4\textwidth,angle=270,origin=c]{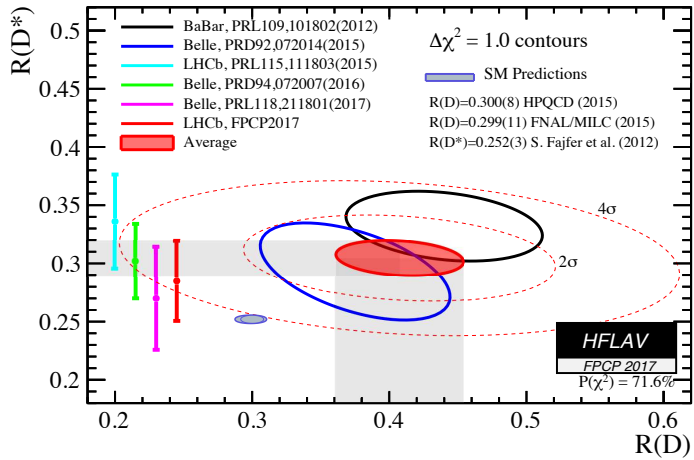}
        \vspace{-2.1cm}
	\end{center}
	\caption{Measurements and SM predictions of $R_\tau(D)$ and
    $R_\tau(D^*)$. Figure taken from the online update
    of~\cite{Amhis:2016xyh}.}
	\label{RDRDs}
\end{figure*}

    \vspace{-3mm}
    \begin{flushright}\textit{Speaker: Florian Bernlochner}\\Karlsruhe Institute of Technology\end{flushright}
    \vspace{-3mm}
    \nopagebreak
The SM predictions for the ratio $R_\tau(D^{(*)})$ are $R_\tau(D)_{\rm SM}
= 0.299 \pm 0.003$ and  $R_\tau(D^*)_{\rm SM} = 0.257 \pm
0.003$~\cite{Bernlochner:2017jka}. The experimental results are
summarised in Figure~\ref{RDRDs}, where the current world average is
shown, as well as the SM predictions.

The BaBar measurements of $R_\tau(D^{})$ and
$R_\tau(D^{(*)})$~\cite{Lees:2012xj,Lees:2013uzd} using hadronic and
leptonic tags for the tau decays show the biggest deviations from the
SM predictions of $\nsigma2$ and $\nsigma{2.7}$, respectively. The Belle
measurements of
$R_\tau(D^{(*)})$~\cite{Huschle:2015rga,Sato:2016svk,Hirose:2016wfn}
are performed with various techniques: with the $\tau$ decaying to
$e\nu\bar\nu$ or $\mu\nu\bar\nu$ and a hadronic and semileptonic tag.
The deviations from the SM are less pronounced, i.e.  below the
$\nsigma2$ level, but consistent with the values provided by BaBar.
Furthermore, \cite{Sato:2016svk,Hirose:2016wfn} contain the
first polarisation measurements in this context.

LHCb measures $R_\tau(D^*)$ from two different decay channels. In
$B_d\rightarrow D^* \tau^- \bar\nu$ with the subsequent decay $\tau
\rightarrow \mu\nu\bar\nu$~\cite{Aaij:2015yra}, finding compatibility
with BaBar and Belle and a $\nsigma{2.1}$ deviation from SM. In $B_d
\rightarrow D^{*-} \tau^+\bar\nu$ with the subsequent decay $\tau
\rightarrow \pi\pi\pi (\pi^0) \nu$ ~\cite{Aaij:2017deq} the result is
only slightly higher than the SM value.

Recently, LHCb has also performed an analysis of the ratio
$R_\tau(J/\psi)$~\cite{Aaij:2017tyk}, obtaining the result 
\begin{align}
R_\tau(J/\psi) = 0.71 \pm 0.17 \pm 0.18\,,
\end{align}
which is significantly above the SM prediction of $[0.25,0.28]$, but
the errors are still large.

Looking into the future, the expected precision for $R_\tau(D^{(*)})$
with a luminosity of $\invfb{10}$ is 4$\%$, and will improve further
to 2$\%$ with a luminosity of $\invfb{22}$. The expected precision
for $R_\tau(D)$ ($R_\tau(D^{*})$) at Belle II is 5.6$\%$ (3.9$\%$)
with $\invab5$ and 3.2$\%$ (2.2$\%$) with $\invab{40}$.

    \needspace{5cm}
    \subsection{\texorpdfstring{$R_\tau(D)$ and $R_\tau(D^*)$}{R(D) and R(D*)} EFT analysis}
    \vspace{-3mm}
    \begin{flushright}\textit{Speaker: Diptimoy Ghosh}\\ICTP Trieste\end{flushright}
    \vspace{-3mm}
    \nopagebreak The combined significance for NP in $b\to c\tau\nu$ processes is at
the $\nsigma4$ level. In an effective field theoretical approach, one
can examine which operators are capable of explaining $R_\tau(D)$ and
$R_\tau(D^*)$ without violating bounds from other
observables~\cite{Bardhan:2016uhr}. Here, three classes of operators
are possible at the dimension-$6$ level: vector, scalar and tensor
operators. While scalar operators cannot explain $R_\tau(D)$ and
$R_\tau(D^*)$ at the same time without violating bounds from the $B_c$
lifetime~\cite{Li:2016vvp,Celis:2016azn,Alonso:2016oyd,Akeroyd:2017mhr}
and $q^2$
distributions~\cite{Freytsis:2015qca,Celis:2016azn,Ivanov:2017mrj},
vector operators provide a good fit to data. In particular, purely
left-handed operators are interesting possibilities. They simply
rescale the SM contributions, leaving $q^2$ distributions unchanged,
and predict $R_\tau(X)/R_\tau(X)_{\rm SM}$ for $X\in\{D,D^*,J/\psi\}$.
However, also tensor operators~\cite{Bardhan:2017xcc} (possibly in
combination with scalar ones) are capable of addressing $b\to
c\tau\nu$ data coherently.

    \needspace{5cm}
    \subsection{Anomalies in \texorpdfstring{$B$}{B} decays: high-\texorpdfstring{$p_T$}{pT} frontier}
    \vspace{-3mm}
    \begin{flushright}\textit{Speaker: Admir Greljo}\\JGU Mainz\end{flushright}
    \vspace{-3mm}
    \nopagebreak\begin{figure*}[t]
	\begin{center}
		\begin{tabular}{cp{7mm}c}
			\includegraphics[width=0.6\textwidth]{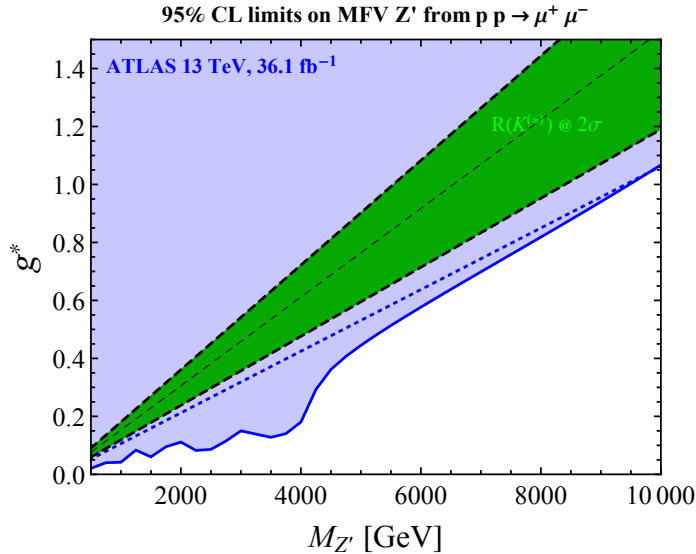}
		\end{tabular}
	\end{center}
	\caption{Limits on the $Z'$ MFV models from $p p \to \mu \mu$. See~\cite{Greljo:2017vvb} for more details.}  
	\label{fig:admir}
\end{figure*}

Anomalies in $B$ meson decays point to a new mass scale potentially
interesting for the ongoing experiments at the high-$p_T$ frontier
(ATLAS and CMS). More precisely, the charged current anomaly ($b \to c
\tau \nu$) implies the effective scale of $\mathcal{O}(\TeV{1})$, while
the neutral current anomaly ($b \to s \mu \mu$) implies the effective
scale of $\mathcal{O}(\TeV{30})$~\cite{DiLuzio:2017chi}. This scale
corresponds roughly to the mediator mass with $\mathcal{O}$(1)
couplings when the effect is tree-level generated. However, some
explicit models exhibit parametric suppression (e.g. from flavour), or
dynamical suppression (e.g. loop-generated), lowering the new mass
scale towards the interesting range for the LHC.

All tree-level models have either colour-neutral vectors ($W^\prime$
and $Z^\prime$) or colour-triplet scalar or vector LQ, or a
combination of those. The expected high-$p_T$ signatures of $Z^\prime$
models are dilepton resonances or non-resonant deviations in the tails
(see e.g.~\cite{Faroughy:2016osc,Greljo:2017vvb}). Interestingly
enough, even if the $Z^\prime$ boson mass is beyond the kinematical
reach for on-shell production, a correlated signal in the tails could
be observed. As shown in Figure~\ref{fig:admir}, the present dimuon
LHC data already sets stringent limits on the $Z^\prime$ models with
minimal flavour violation (MFV) explaining $b \to s \mu
\mu$~\cite{Greljo:2017vvb}. 

The $B$ physics anomalies inspired LQ searches at the LHC exhibit
interesting interplay between three production mechanisms: LQ pair
production, single LQ production in association with the lepton, and
di-lepton production (see e.g.~\cite{Dorsner:2018ynv} for a recent
discussion). In particular, as shown in~\cite{Faroughy:2016osc},
the present ditau LHC data sets stringent limits on most models for $b
\to c \tau \nu$ with exclusive coupling to the third family. Sizeable
bottom-strange flavour violation can partially relax these
constraints~\cite{Buttazzo:2017ixm,Crivellin:2017zlb,Dorsner:2017ufx},
introducing potentially dangerous flavour changing neutral currents in
the down quark sector.

The combined explanation of $B$ anomalies singles out the $U_1$ vector
leptoquark representation as a viable single mediator
model~\cite{Buttazzo:2017ixm}. As shown in~\cite{Buttazzo:2017ixm},
going beyond LHC is required to fully explore the relevant parameter
space. However, when considering an explicit ultraviolet completion of
this simplified model
(e.g.~\cite{DiLuzio:2017vat,Bordone:2017bld,Barbieri:2017tuq,Greljo:2018tuh}),
a coloron decaying to dijet is predicted to be around the corner at
the LHC.

    \needspace{5cm}
    \subsection{Simultaneous explanations of \texorpdfstring{$R_\tau(D)$, $R_\tau(D^*)$ and $b\to s\mu^+\mu^-$}{R(D), R(D*) and b->s mu mu} data}
    \vspace{-3mm}
    \begin{flushright}\textit{Speaker: Gino Isidori}\\University of Zurich\end{flushright}
    \vspace{-3mm}
    \nopagebreak\begin{figure*}[t]
	\begin{center}
        \includegraphics[width=0.6\textwidth]{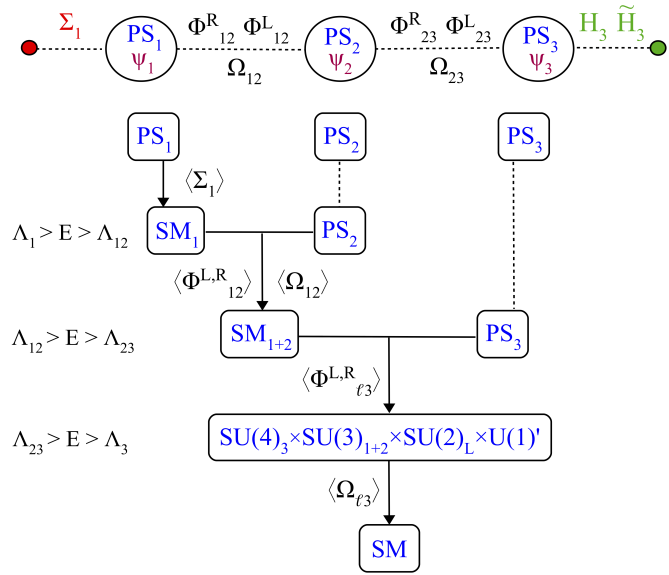}
	\end{center}
	\caption{Outline of the model  ${\rm PS}_{1}\times {\rm
    PS}_{2}\times {\rm PS}_{3}\to {\rm SM}$, that provides an explanation for both flavour anomalies and 
    Yukawa hierarchies. The high scale ${\rm PS}_{1}$ breaking lies at
    $\Lambda_1\sim \TeV{10^3}$, while ${\rm SM}_{1+2}\times {\rm
    PS}_{3}\to {\rm SM}$ occurs in two steps at lower energies
    ($\Lambda_{23}\sim \TeV{20}$
    and $\Lambda_{3}\sim \TeV{1}$).  Figure taken from \cite{Bordone:2017bld}.}  
	\label{fig:SSBdiag}
\end{figure*}
The anomalies are only seen in semi-leptonic processes and indicate
non-vanishing coefficients of left-handed vector operators. In order
to explain $R_\tau(D^{(*)})$, NP of $\mathcal{O}(10\%)$ of the
tree-level SM contribution is required. This is in contrast to $b\to
s\mu^{+}\mu^{-}$ which is a loop- and CKM-suppressed process in the SM
and therefore NP of approximately 25\% of this (suppressed)
contribution is required.

From the phenomenological point of view, this indicates NP with
$\mathcal{O}(1)$ couplings to third generation, moderate couplings to
the second generation and small or vanishing couplings to the first
generation, i.e.~a structure similar to the one appearing in the SM 
Yukawa couplings.

One can follow two approaches to explain both anomalies simultaneously.
First, using an EFT approach, it can be shown that both anomalies can be
explained simultaneously without any fine tuning~\cite{Buttazzo:2017ixm}.
The corresponding EFT solution is not unique, but its main features are stable
once we require a fit to both anomalies. In particular, left-handed four-fermion 
operators with almost equal strength for singlet and triplet terms 
are needed.

Second, one can try to derive these effective operators using
simplified dynamical models.  Among the possible options, the vector
LQ singlet $U_{1}$ with quantum numbers $U_{1}\sim(\bar{3},1,-4/3)$
gives a particularly good fit to data (with no fine
tuning)~\cite{Barbieri:2015yvd}.  However, the introduction of a
massive vector LQ requires a Higgs sector in order to achieve an UV
complete model. Recently, several models that use an $SU(4)$
Pati-Salam symmetry and give renormalisable massive vector LQ were
proposed~\cite{DiLuzio:2017vat,Calibbi:2017qbu,Bordone:2017bld,Barbieri:2017tuq,Blanke:2018sro}.
In~\cite{Bordone:2017bld} a model in which at high energies the three
families are charged under three different gauge groups ${\rm
PS}_{i}={\rm SU}(4)_{i}\times [{\rm SU}(2)_{L}]_{i}\times [{\rm
SU}(2)_{R}]_{i}$ is introduced (see Figure~\ref{fig:SSBdiag}). The
interest of this model is that it offers not only a solution to the
anomalies, but also an explanation for the observed hierarchies in the
Yukawa couplings of quarks and leptons.

    \needspace{5cm}
    \subsection{Experimental status and prospects for \texorpdfstring{$\mu\to e$}{mu->e} experiments}
    \vspace{-3mm}
    \begin{flushright}\textit{Speaker: Angela Papa}\\Paul Scherrer Institute\end{flushright}
    \vspace{-3mm}
    \nopagebreak

Some of the most severe bounds of LFV come from $\mu\to e$ experiments
because these processes are practically zero in the SM. For example,
the neutrino oscillation induced $\mu^+\to e^+\gamma$ branching ratio
lies with $10^{-54}$ far below the experimental reach.  The other
`golden channels' for $\mu\to e$ flavour violation are $\mu\to eee$
and $\mu N\to e N$. A summary of the current and expected future
bounds for these processes is shown in Table~\ref{tab:muebounds}.

\begin{figure}[b]
\centering
\begin{tabular}{c|c|c}
&Current upper limit & Future sensitivity\\\hline
$\mu  \to e\gamma$ & $4.2\times 10^{-13}$ \cite{TheMEG:2016wtm}  &$4\times 10^{-14}$ (MEG II~\cite{Baldini:2013ke,Baldini:2018nnn})\\
$\mu  \to eee    $ & $1.0\times 10^{-12}$ \cite{Bellgardt:1987du}&$1\times 10^{-16}$ (Mu3e~\cite{Blondel:2013ia, Perrevoort:2016nuv, Bachmann:2014bfa})\\
$\mu N\to e N    $ & $7.0\times 10^{-13}$ \cite{Bertl:2006up}    &$\le     10^{-16}$ (Mu2e~\cite{Carey:2008zz, Kutschke:2011ux} and COMET~\cite{Cui:2009zz})
\end{tabular}
\renewcommand{\figurename}{Table}
\caption{Current and future bounds for $\mu\to e$ processes}
\label{tab:muebounds}
\end{figure}

Because $\mu\to e\gamma$ and $\mu\to eee$ are coincidence experiments,
they benefit from a continous (DC) muon beam like the one at PSI with
an intensity of $\mathcal{O}(10^8) \mu/\mathrm{s}$. On the other hand,
non-coincidence experiments measuring $\mu-e$ conversion need a pulsed
beam that reaches $\mathcal{O}(10^{11})\mu/\mathrm{s}$. Because the
future MEG II and Mu3e experiments have similar beam requirements
(high intensity DC-beam at low momentum) they will share the PiE5 beam
line of the HIPA accelerator at PSI. There are efforts ongoing to
increase the intensity by two more orders of magnitude by 2025.

The clean back-to-back signature of $\mu\to e\gamma$ in MEG can be
separated from the radiative muon decay $\mu\to\nu\bar\nu e\gamma$ or
accidental backgrounds $\mu+\gamma\to\nu\bar\nu e+\gamma$. MEG II will
double the resolution and use twice the beam intensity of MEG to
further push the bound. MEG II will conduct a full engineering run in 2018.

For Mu3e the signature is more complicated and requires a very good
energy resolution to fully reconstruct $\sum E_i=m_\mu$. This controls
the background from the rare muon decay $\mu\to\nu\bar\nu eee$. To
control accidental backgrounds, high timing and position resolution is
necessary. To that end, the pixel layers' thickness does not exceed
$0.1\%$ of the radiation lengths. In contrast to MEG, Mu3e will use
the full rate of $10^8\mu/\mathrm{s}$. A pre-engineering run is
expected in 2020.

$\mu-e$ conversion experiments such as Mu2e and COMET search for a
single mono-energetic electron. For that, they require a very large
number of stopped muons $\mathcal{O}(10^{18})$. Both experiments need
a good energy resolution to reject backgrounds from a muon decay in
orbit. Further, an extinction factor of the beam of
$\mathcal{O}(10^{-10})$ and precise timing is required to handle beam
related backgrounds. This is further aided by having the stopping
target and the production target well separated. The third major
background are cosmic rays that may fake a high energy electron in the
detector. The first phase of the COMET experiment is expected to have
a full engineering run 2018 while Mu2e is still under construction.

\vspace{-1cm}

    \needspace{5cm}
    \subsection{Connections between \texorpdfstring{$R_\mu(K)$, $R_\mu(K^*)$ and $\mu\to e$}{R(K), R(K*) and mu->e} processes}
    \vspace{-3mm}
    \begin{flushright}\textit{Speaker: Ivo de Medeiros Varzielas}\\Instituto Superior Tecnico\end{flushright}
    \vspace{-3mm}
    \nopagebreak
An explanation of $R_\mu(K)$ and $R_\mu(K^{*})$ requires new physics
couplings to muons and/or electrons. Among the possible extensions of
the SM such as $Z^{\prime}$ models, compositeness or SUSY, this talk
focuses on LQ. As shown in a model independent analysis, LQ can
provide a solution to these anomalies \cite{Hiller:2014yaa,
Hiller:2016kry, Hiller:2017bzc}. Based on~\cite{Varzielas:2015iva} the
impact of LQ on the LFV processes $B\to K\ell^{\prime}\ell$,
$B\to\ell^{\prime}\ell$ and $\ell\to\ell^{\prime}\gamma$ has been
analysed, explaining $R_\mu(K)$ and $R_\mu(K^*)$. The focus of this
analysis were the scalar triplet with quantum numbers
$\Delta\sim(3,3,-2/3)$ and scalar doublet the $\Delta\sim(3,2,-1/3)$.
Experimental limits such as $\mu\to e\gamma$,
$B_{s}-\bar{B}_{s}$-mixing and $B\to K\mu e$ give constraints on the
parameter space and a combined analysis leads to an upper limit of the
LQ mass of $\TeV{50}$, which could be within the reach of future
colliders. The constraints will become more stringent in the coming
years, see Table~\ref{tab:muebounds}.

Experimental data favour a Yukawa-like hierarchical structure of the
LQ couplings. From the theoretical point of view this can be achieved
in a natural way e.g. introducing an additional ${\rm SU}(3)_{F}$
symmetry for SM fermions which is then broken by the vacuum
expectation value of the so-called familon. This way, one can either
have or avoid LQ coupling to electrons and hence, have models that may
be tested in the future through important constraints, e.g.  $\mu\to
e\gamma$~\cite{Crivellin:2017dsk}.

    \needspace{5cm}
    \subsection{Correlations with kaon physics}
    \vspace{-3mm}
    \begin{flushright}\textit{Speaker: Martin Hoferichter}\\University of Washington\end{flushright}
    \vspace{-3mm}
    \nopagebreak The flavour anomalies discussed so far suggest to search for LFV and
LFUV in the kaon sector~\cite{Crivellin:2016vjc}. Here the decays
$K\to\pi\ell^+\ell^-$ and $K\to \ell^+\ell^-$ are especially
interesting but also very challenging: long-distance contributions
from the SM need to be separated from the interesting short-distance
effects, both of which enter in poorly known low-energy constants of
the expansion in chiral perturbation theory.
However, in the context of LFUV, this complication is absent if the
difference between electron and muon parameters is considered. This
simplification is due to the fact that in the SM all interactions
(except for the Higgs boson Yukawa couplings) are LFU conserving.
Since the  Higgs corrections are negligible, it follows that the SM
decays of kaons to muons or electrons differ only by phase-space
factors. Thus, any deviation from the SM predictions must be related
to LFUV NP which is necessarily short-distance once the new particles
are assumed to be heavy. Assuming MFV, the derived limits from kaon
decays would need to be improved by at least an order of magnitude in
order to probe the parameter space relevant for the explanation of the
$B$ meson anomalies and thereby test those anomalies within the MFV
hypothesis. However, it is well possible that NP does not respect MFV,
so that effects in kaon decays could be observed earlier, or, if not,
non-MFV models that predict so would be excluded.

Another anomaly that could in principle be related to $b\to
s\mu^+\mu^-$ and $b\to c\tau\nu$ is the $CP$ asymmetry in $\tau\to
K_s\pi\nu$ which, as measured by the BaBar collaboration, differs from
the SM prediction by $\nsigma{2.8}$~\cite{BABAR:2011aa}. Most
non-standard interactions do not allow for the required strong phase
needed to produce a non-vanishing $CP$ asymmetry, leaving only new
tensor interactions as a possible
mechanism~\cite{Devi:2013gya,Dhargyal:2016jgo}.  However, contrary to
previous assumptions in the literature, the crucial interference
between vector and tensor phases is suppressed by at least two orders
of magnitude due to Watson's final-state-interaction
theorem~\cite{Watson:1954uc}.  Furthermore, the strength of the
relevant $CP$-violating tensor interaction is strongly constrained by
bounds from the neutron electric dipole moment~\cite{Afach:2015sja}
and $D$--$\bar{D}$ mixing.  These observations together imply that it
is extremely difficult to explain the current $\tau\to K_s\pi\nu_\tau$
measurement in terms of physics beyond the SM originating in the
ultraviolet regime~\cite{Cirigliano:2017tqn}.

\vspace{-1cm}

    \needspace{5cm}
    \subsection{Summary and outlook}
    \vspace{-3mm}
    \begin{flushright}\textit{Speaker: Andreas Crivellin}\\Paul Scherrer Institute\end{flushright}
    \vspace{-3mm}
    \nopagebreak The experimental hints for lepton flavour universality violation are
convincing since they form a consistent picture: large effects related
to tau leptons, moderate effects related to muons while electron
channels seem to be SM-like. Confirming these hints would probably be
the biggest breakthrough in particle physics since the discovery of
parity violation in 1956. Accounting for LFUV would require some
radical NP, providing us with a convincing physics case for a future
collider, and would lead us to a golden age for particle physics. On
the other hand, if these hints for NP disappeared, it would be well
possible that we are ahead of a desert ranging over many orders of
magnitude in energy. However, if one believes in statistics and that
the experimental analysis as well as the theory predictions were done
correctly, the first option is fortunately much more likely.

\section{Conclusion}

\begin{figure*}[t]
	\begin{center}
			\includegraphics[width=0.7\textwidth]{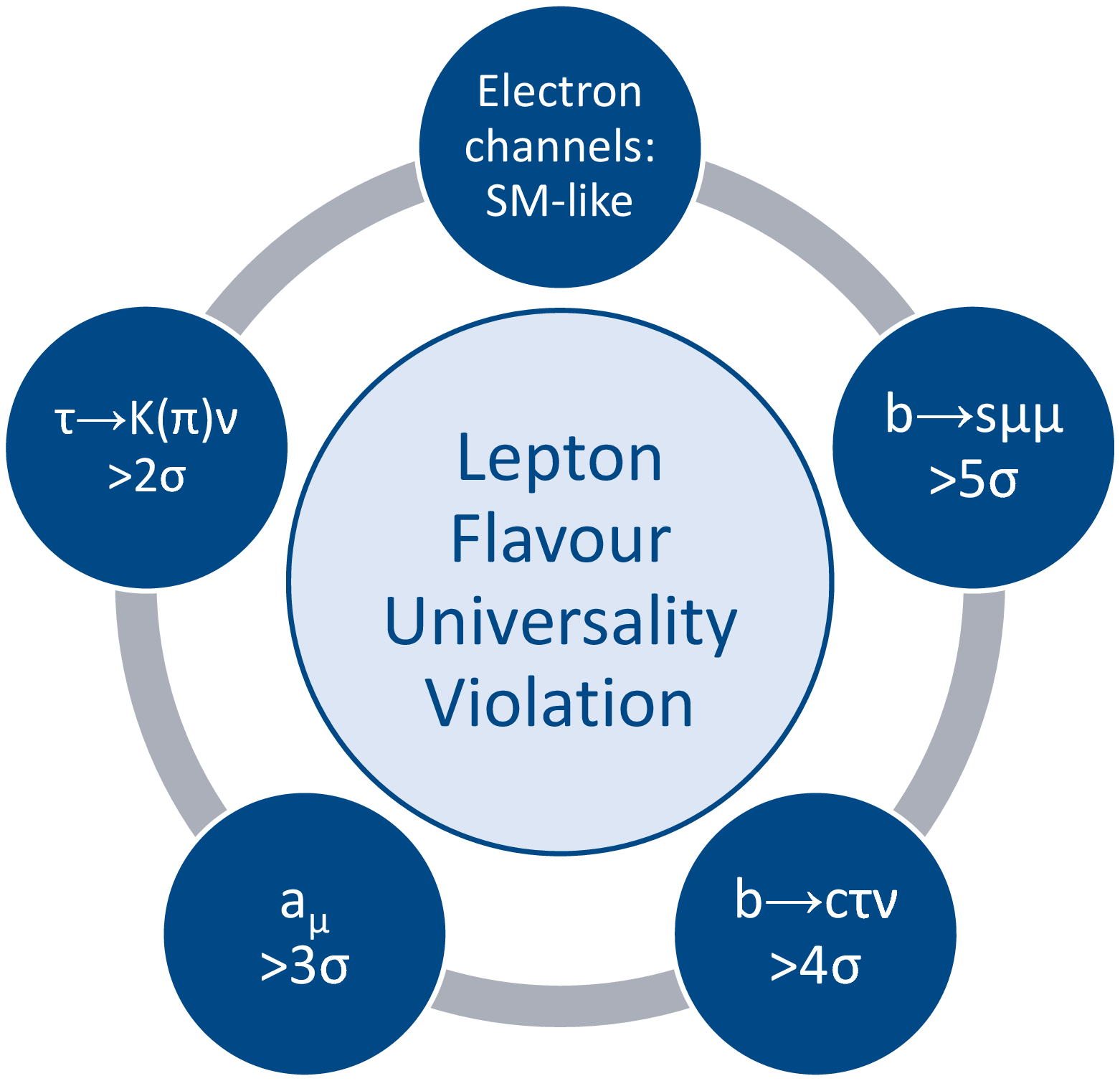}
	\end{center}
	\caption{Deviations from the SM predictions pointing at the
    violation of lepton flavour universality. In addition to the hints
    discussed in this workshop, we included the anomalous magnetic
    moment of the muon~(see e.g. \cite{Nyffeler:2016gnb}) and $V_{us}$
    extracted from tau decays~\cite{Lusiani:2017spn}.}         
	\label{fig:snf}
\end{figure*}

In recent years we accumulated intriguing hints for physics beyond the
SM in $b\to c\tau\nu$ and $b\to s\mu^+\mu^-$ processes. These signs
of NP are accompanied by more hints for NP whose common pattern is the
violation of lepton flavour universality (see
Figure~\ref{fig:snf}).

In this workshop we first considered the experimental and theoretical
status of $b\to s\mu^+\mu^-$ transitions with focus on
$B_s\to\ell^+\ell^-$. Currently, the deviations from the SM predictions
are dominated by the measurements of $R_\mu(K^{(*)})$ and
$P_5^\prime$.  However, because of its theoretical cleanliness, in the
future $B_s\to\mu^+\mu^-$ will play a crucial role distinguishing
various NP scenarios, i.e. confirming or disproving whether NP couples
also axial-vectorially (in addition to vectorial couplings) to muons
or not.

The second day of the workshop began with the study of the anomalies
in $b\to c\tau\nu$ processes and its connections to $b\to
s\ell^+\ell^-$ transitions. While a common origin of the two anomalies
seems plausible, simultaneous explanations are challenging and further
data will decide whether the discrepancies in $b\to c\tau\nu$ persist.
Finally, we discussed the implications of $R_\mu(K^{(*)})$ for
experiments searching for $\mu\to e$ flavour violation and LFV in kaon
decays.

Furthermore, (even though we did not discuss it in detail in this
workshop) the anomalous magnetic moment of the muon can be linked to
the hints for LFUV in $B$ decays as well and $\tau$ decays provide
additional interesting tests of the SM.  

\acknowledgments{We gratefully acknowledge support of the University
of Zurich (UZH) and the Paul Scherrer Institute and the Swiss
National Science Foundation's \textit{Scientific Exchanges} program
under contract 177983. A special thanks to A.~Van Loon-Govaerts for
her help in organising the workshop. 
A.~Crivellin and D.~M\"uller are supported by an Ambizione Grant of
the Swiss National Science Foundation (PZ00P2\_154834). Y.~Ulrich and
M.~Ghezzi are supported by the Swiss National Science Foundation under
contract 200021\_163466 and 200021\_160156, respectively.
S.~Descotes-Genon acknowledges partial support from Contract
FPA2014-61478-EXP and from the European Union's Horizon 2020 programme
under grant agreements No 690575, No 674896 and No. 692194.
M.~Hoferichter acknowledges support from US DOE Grant No.
DE-FG02-00ER41132.
M.~Misiak acknowledges partial support by NCN (Poland) under research
projects 2017/25/B/ST2/00191 and UMO-2015/18/M/ST2/00518.

}

\bibliographystyle{JHEP}
\bibliography{BIB}{}

\end{document}